\documentclass[11pt]{article}
\usepackage{amsthm,amsfonts,amsmath,amssymb}
\usepackage{graphicx}
\theoremstyle{plain}

\newtheorem{thm}{Theorem}[section]

\theoremstyle{definition}

\newtheorem{rem}[thm]{Remark}
\theoremstyle{remark}

\def\jring{j^{\rm ring}}
\def\Pring{P^{\rm ring}}
\def\Qring{Q^{\rm ring}}
\def\jsemi{j^{\text{s-i}}}
\def\rsemi{r^{\text{s-i}}}
\def\Qsemi{Q^{\text{s-i}}}
\def\Gammahat{\hat \Gamma}
\def\cring{c^{\rm ring}}

\newcommand{\be}{\begin{equation}}
\newcommand{\ee}{\end{equation}}
\newcommand{\bea}{\begin{eqnarray}}
\newcommand{\eea}{\end{eqnarray}}

\newcommand{\cthm}[1]{Theorem~\ref{#1}}

\newcommand{\cfig}[1]{Figure~\ref{#1}}
\newcommand{\crem}[1]{Remark~\ref{#1}}

\newcommand{\E}[1]{\bE{\left[#1\right]}}

\newcommand\bE{{\mathbb E}}

\newcommand\bR{{\mathbb R}}

\newcommand\bZ{{\mathbb Z}}
\newcommand\bbz{{\mathbb Z}}

\numberwithin{equation}{section} 
\title{\null\vskip-60pt 
  The blockage problem\footnote{Dedicated to Raghu Varadhan.  This
    problem was part of a talk by JLL presented at the celebration of
      Raghu's seventieth birthday.  We hope that this note will prompt
      Raghu to solve the problem.}}
\date{\today}
\author{{O. Costin${}^1$, J. L. Lebowitz${}^{2,3}$, E. R. Speer${}^2$, and
    A. Troiani${}^4$ }\\ \\
{\small $^1$ Department of Mathematics, The Ohio State University,}\\
{\small 231 W. $18^{\rm th}$ Avenue, Columbus, OH 43210 USA}\\
{\small $^2$ Department of Mathematics, Rutgers University,}\\
{\small Piscataway NJ 08854-8019 USA}\\
{\small $^3$ Department of Physics, Rutgers University,}\\
{\small Piscataway NJ 08854-8019 USA}\\
{\small $^4$ Mathematical Institute, Leiden University,}\\
{\small P.O. Box 9512, 2300 RA Leiden, The Netherlands}}

\begin{document}

\maketitle

\begin{abstract} We investigate the totally asymmetric exclusion process on
$\bbz$, with the jump rate at site $i$ given by $r_i=1$ for $i\ne0$,
$r_0=r$.  It is easy to see that the maximal stationary current $j(r)$ is
nondecreasing in $r$ and that $j(r)=1/4$ for $r\ge1$; it is a long
outstanding problem to determine whether or not the critical value $r_c$ of
$r$ such that $j(r)=1/4$ for $r>r_c$ is strictly less than 1.  Here we
present a heuristic argument, based on the analysis of the first sixteen
terms in a formal power series expansion of $j(r)$ obtained from finite
volume systems, that $r_c=1$ and that for $r\lessapprox1$,
$j(r)\simeq{1/4}-\gamma\exp[-{a/(1-r)}]$ with $a\approx2$.  We also give
some new exact results about this system; in particular we prove that
$j(r)=J_{\rm max}(r)$, with $J_{\rm max}(r)$ the hydrodynamic maximal
current defined by Sepp\"al\"ainen, and thus establish continuity of
$j(r)$.  Finally we describe a related exactly solvable model, a
semi-infinite system in which the site $i=0$ is always occupied.  For that
system, $\rsemi_c=1/2$ and the analogue $\jsemi(r)$ of $j(r)$ satisfies
$\jsemi(r)=r(1-r)$ for $r\le\rsemi_c$;  $\jsemi(r)$ is the limit
of finite volume currents inside the curve $|r(1-r)|=1/4$ in the complex
$r$ plane and suggest that analogous behavior may hold for the original
system.  \end{abstract}

\section{Introduction}\label{sec:intro}

Our understanding of nonequilibrium stationary states (NESS) of macroscopic
systems is very incomplete at present.  While systems in equilibrium
exhibit behavior which, away from phase transitions, is essentially local,
the behavior in a NESS is global.  That is, away from criticality, changes
made in one region of an equilibrium system have almost no effect far away,
but any local change in systems carrying fluxes of conserved quantities
will almost certainly be felt at a distance.

This difference in behavior is particularly striking in low dimensional
systems, where local changes in conductivity can radically affect the total
flux through the system.  Some years ago one of us (JLL), together with S.
Janowsky, studied a very simple system of this kind, a perturbation of the
familiar one-dimensional totally asymmetric simple exclusion process
(TASEP).  Recall \cite{Ligg,Ligg2} that the latter is a model in which
particles occupy some of the sites of a one dimensional lattice; we let
$\eta_i=1$ if a particle is present on site $i$, $\eta_i=0$ otherwise. The
particles independently try to jump to the site on their right with rate
one, succeeding if the target site is empty and otherwise staying at their
original site.  The lattice may be either (i)~a ring, (ii)~a finite
interval with open boundaries which particles enter at rate $\alpha$ on the
left and exit at rate $\beta$ on the right, or (iii)~all of ${\bZ}$.  The
invariant measures for these systems are known: (i)~on a ring with $L$
sites and $N$ particles all $\binom L N$ configurations have equal
probability; (ii)~on the interval the system has a rich phase diagram in
the $\alpha$-$\beta$ plane, obtained by the ``matrix ansatz'' \cite{DEHP}
or otherwise \cite{SD}; (iii)~on ${\bZ}$ \cite{Ligg} there are translation
invariant measures, the extremal ones of which are product measures
$\nu_\rho$, $0\le\rho\le1$, in which the probability of a site being
occupied is $\rho$.  In all cases there is a steady state current given by
$\langle\eta_i(1-\eta_{i+1})\rangle$ (the value is independent of $i$),
where for any function $f(\eta)$ of the configuration we write
$\langle f\rangle$ for the expectation of $f$ in the stationary state.  The
current in the state $\nu_\rho$ on $\bZ$ is thus $\rho(1 - \rho)$, with
maximum value 1/4 occurring for $\rho = 1/2$.  (On $\bZ$ there are also
so-called ``blocking'' measures with zero current, which will play no role
in our discussion.)

Janowsky and Lebowitz \cite{JL1,JL2} studied what happens in these systems
under the introduction of a {\em blockage}: a change in the jump rate, from
1 to some value $r\ge0$, across one bond of the system.  Consider first the
system on $\bZ$, and suppose without loss of generality that the blockage
is on the bond between site 0 and site 1.  We will be interested in the
stationary states of the system and in particular in $j(r)$, the maximum
value, taken over all stationary states, that the current can assume.
Clearly if $r = 0$ then $j(r)=0$ since all stationary states will have zero
current: any particles in the system will just pile up behind and empty out
in front of the blockage.  For $r \geq 1$ the ``blockage'' in fact offers
no impediment, in the sense that $j(r)=1/4$ (see discussion below); more
generally, one expects that in this case the blockage will have only a
local effect and that, just as in the original ($r=1$) system, there will
be for any $\rho\in[0,1]$ a stationary state with current $\rho(1-\rho)$ in
which the measure far from the blockage approaches $\nu_\rho$, and thus in
particular that $j(r)=1/4$.  The question then is what is the nature of the
stationary measures for $r \in (0,1)$?  How will the system respond
globally to this local change?  It may be convenient to think in terms of
traffic flow in which the width of one stretch of the road is changed.  How
well can the traffic adjust to a narrowing of the road?

A simple mean field approximation  gives a current $r/(1+r)^2$ for
$r\le1$, corresponding to $r_c=1$ \cite{JL2}.  Beyond this, some precise
information can be obtained by considering the system on $\bZ$ as the
infinite volume limit of the system either on the ring or on an interval.
Consider in particular the open system on an interval of $2L$ sites labeled
$-L+1,\ldots,L$, with the special bond between sites $0$ and $1$ and with
$\alpha,\beta\ge1$.  It can be shown by a coupling argument \cite{JL2} that
the current $j^{\alpha,\beta}_L(r)$ in this system is decreasing in $L$ for
fixed $r$ and, by a similar argument, that $j^{\alpha,\beta}_L(r)\ge j(r)$.
A limiting NESS exists as $L\to\infty$ (in theory one may need to take this
limit along a subsequence, but that has no effect on the conclusions we
draw below, and we ignore this possibility for notational convenience) so
that
 \be\label{currlim}
j(r)=\lim_{L\to\infty}j^{\alpha,\beta}_L(r), \qquad \alpha,\beta\ge1,
 \ee
  (no subsequence is needed here).  Since clearly $j^{\alpha,\beta}_L(r)$
is nondecreasing in $r$ at fixed $L$, $j(r)$ must also be nondecreasing in
$r$.  Moreover, $j(r)\le1/4$ for all $r$, as one may see by an argument
based on \cite[p.~273]{Ligg2}: consider an unbounded sequence of translates
of the stationary state for which a limiting state exists; this will be a
steady state for the unperturbed system on $\bZ$ with all rates equal to 1
and thus have current at most $1/4$.  Since $j(1)=1/4$, there exists a
critical value $r_c\in[0,1]$ with $j(r)<1/4$ for $r<r_c$ and $j(r)=1/4$ for
$r>r_c$.  The key question that we would like to answer is whether or not
$r_c=1$.

Exact formulas for $j_L^{\alpha,\alpha}(r)$ for $L=1,2,3$ and for
$j_4^{1,1}(r)$ were obtained in \cite{JL2}.  These show that
$j_4^{1,1}(0.51)<1/4$, which implies that $r_c>0.51$: this is unfortunately
the best rigorous lower bound that we know, although numerical simulations
strongly support the conclusion that $r_c>0.75$.  Simulations on large open
systems in \cite{JL2} gave a smooth concave curve for
$j_L^{1,1}(r)\approx j(r)$, $r \in [0,1]$ which appeared to approach the
value ${1/4}$ very flatly.  Janowsky and Lebowitz also proposed an analytic
form which fits this data very well:
\be\label{ljf}
 j(r) \approx {1/4} - \exp[- {G(r)/(1-r) H(r)}], \quad \quad r
\in [0,1],
 \ee
 with $G$ and $H$  polynomials determined (as Pad\'e approximants) from
the six terms in the power series $J(r)$ given in \eqref{bjdef} below.

There has been further rigorous work on this and related problems.  Covert
and Rezakhanlou \cite{CR} study the hydrodynamic limit of a system in which
the site-dependent jump rate is a discretization of a uniformly positive,
bounded, and continuously differentiable function $r(x)$, $x\in\bR$, and
prove the existence of a hydrodynamic scaling limit in which the evolution
of the density $\rho(x,t)$ is governed by the Burgers' equation
$\partial_t\rho+ \partial_x[r\rho(1-\rho)]=0$.  Sepp\"al\"ainen \cite{S}
proves the existence of the hydrodynamic limit of the system with a single
slow bond, as considered here.  He shows also that the maximum
current $J_{\rm max}(r)$ in the model, as he defines it, is a continuous
function of $r$, but leaves open the question of whether or not his and our
definitions of the maximal current agree.   We sketch in
Appendix~\ref{cont} an argument establishing that $J_{\rm max}(r)=j(r)$, so
that in particular $j(r)$ is continuous.

As to the value of $r_c$, it is argued in \cite{dN}, from a finite size
scaling analysis of simulation data, that $r_c\approx0.80$.  On the other
hand, our process is directly related to a certain last passage percolation
model \cite{S,Joh,PS}, and this in turn is equivalent to a directed polymer
model on the lattice, at zero temperature, with a line of defects, that is,
with a columnar ``pinning'' potential; see \cite{PS} for a review.  In this
language $r_c=1$ corresponds to the pinning of the polymer for arbitrarily
small values of the potential.  The current interest in such systems, which
have been studied in both the physics and mathematics literature, is one
motivation for the study of the blockage model.  In the physics literature,
renormalization group arguments \cite{HN} indicate that a directed polymer
at low temperature in the continuum is indeed localized for all values of
the potential, giving some support to $r_c=1$.  Mathematically, Beffara et
al.~\cite{BSV} study a model of random polynuclear growth with a columnar
defect and find a phase transition at a nonzero value of the pinning
potential, although one of the authors believes that $r_c=1$ for the
blockage problem \cite{vS}.

 The present paper is primarily concerned with the value of $r_c$ and the
nature of the limit $j(r)\to1/4$ as $r\to r_c$.  We give arguments that
$r_c = 1$ and that for $r\lessapprox1$,
$j(r)\simeq{1/4}-\alpha\exp[-{A/(1-r)}]$ with $A\approx2$, a form
consistent with \eqref{ljf}.  These arguments are far from a proof; they
are based on the analysis of sixteen terms of a formal power series whose
convergence is not known but whose derivation suggests that it may
represent $j(r)$.

 \smallskip\noindent
 {\bf Remark: The system on a ring.} For computations it is frequently
convenient to consider a finite system on a ring rather than on an
interval; we always consider a ring with $2L$ sites, labeled as for the
interval above and again with the special bond joining $0$ and $1$, and
with $L$ particles.  We believe on the basis of the behavior when $r=1$,
the exact solution of small systems, and the simulation of large systems,
that the current $\jring_L(r)$ in this system is decreasing in $L$ and
satisfies $\lim_{L\to\infty }\jring_L(r)=j(r)$, but we do not have a proof.
A coupling argument leads to the inequality
$j^{\alpha,\beta}_L(r)\ge \jring_{L+1}(r)$ for $\alpha,\beta\ge1$.

The rest of the paper is organized as follows.  The derivation of the power
series referred to above is presented in Section~\ref{sec:ps} and its
analysis is given in Section~\ref{sec:current}.   In
Section~\ref{sec:semi} we discuss a simpler model, a semi-infinite TASEP on
sites $1,2,\ldots$ in which particles enter site 1 at rate $r$; here the
relation between the current and its power series can be completely
determined, and we argue that the results shed some light on the properties
of the original system.  Some technical points are relegated to appendices.

\section{Power series \label{sec:ps}}

 The steady state probabilities in any of the finite systems considered
above---on the ring or the interval---are rational functions of $r$, since
they may be obtained as ratios of minors of the generator $M(r)$, which is
affine in $r$ (see \eqref{Mdef}  below).  The current is thus also rational in
$r$; it is nonsingular (and in fact vanishing) at $r=0$, and so has a
power series in $r$ convergent in some neighborhood of the origin:
 \be\label{allseries}
 j_L^{\alpha,\beta}(r)=\sum_{k=0}^\infty  c_{L,k}^{\alpha,\beta}\,r^k,\qquad 
 \jring_L(r)=\sum_{k=0}^\infty  \cring_{L,k}\,r^k.
 \ee
  It is shown in \cite{JL2} for the system on the interval, and may be
verified similarly for the half-filled ring, that for $k\le L$ the
coefficients in \eqref{allseries} agree and are independent of $L$,
$\alpha$, and $\beta$ as long as $\alpha,\beta\ge1$; we denote the common
value as $c_k$:
 \be\label{coeffind}
 c_k=c_{L,k}^{\alpha,\beta}=\cring_{L,k}, \qquad k\le L,\quad \alpha,\beta\ge1.
 \ee
 Thus for $j_L^*=j_L^{\alpha,\beta}$ or $j_L^*=\jring_L$,
 \be\label{jbps}
 j_L^*(r)=\sum_{k=0}^Lc_kr^k+O(r^{L+1}),
 \ee
 where the remainder term, but not the $c_k$, may depend on $L$ and on
 which current is under consideration.

 More generally, one can by similar methods obtain a corresponding result
for the entire measure.  The proof of the next result, and of the extension
mentioned in \crem{th1rmk}(a), is implicit in the proof of Theorem~1 in
Appendix~B of \cite{JL2}.

\begin{thm}\label{gencoeffs} Suppose that $L\ge j\ge1$, and let $f(\eta)$
be a function of the configuration which depends only on the $2j$ variables
$\eta_{-j+1},\ldots\eta_j$.  Then in each of the finite systems we are
considering $\langle f\rangle$ has a power series in $r$ for which the
coefficients up to order $L-j$ are independent of $L$.  Moreover, on the
interval these coefficients are independent of $\alpha$ and $\beta$, for
$\alpha,\beta\ge1$, and are the same  as on the ring.\end{thm}

\begin{rem}\label{th1rmk} (a) In the special case of an open system with
$\alpha=\beta=1$ one in fact obtains one more order, that is, the
coefficients of the power series of $\langle f\rangle$ up to order $L-j+1$
are independent of $L$.

 \smallskip \noindent
 (b) \eqref{coeffind} is a corollary of the $j=1$ case of
\cthm{gencoeffs}, since the current in any of these systems has value
$r\,\langle\eta_0\,(1-\eta_1)\rangle$.\end{rem}

We may compute the power series \eqref{allseries}, or indeed the series
for any $\langle f \rangle$ on the ring or interval, by a direct expansion,
without obtaining the exact rational form (which is usually a much more
computationally intensive problem).  For if the steady state measure
assigns probability $P(\eta)=\sum_{k=0}^\infty p_k(\eta)\,r^k$ to the
configuration $\eta$ then $P$ is the null vector of the generator $M(r)$ of
the process:
 \be\label{nullv}
\sum_\eta M(r)_{\zeta\eta}P(\eta)=0,\qquad \forall\zeta.  
 \ee
 $M(r)$ has the form $M(r)=M^{(0)}+rM^{(1)}$ and $p_0(\eta)$ is known, so
that $p_1(\eta)$, $p_2(\eta)$,
etc.~may be found recursively by solving  
 \be\label{tosolve}
   M^{(0)}p_k = -M^{(1)}p_{k-1}.
 \ee
 For example, for the ring $p_0(\eta)=\delta_{\eta,\eta^{(0)}}$ with
$\eta^{(0)}_i=1$ if $i\le0$ and $\eta^{(0)}_i=0$ if $i\ge1$, and $M$ is a
$\binom{2L}{L}\times\binom{2L}{L}$ matrix with
 \be\label{Mdef}
M_{\zeta\eta}=\begin{cases}
   \eta_i(1-\eta_{i+1}),&\hbox{if $\zeta=\eta^{i,i+1}$, $i\ne0$,}\\
   \eta_0(1-\eta_1)r,&\hbox{if $\zeta=\eta^{0,1}$,}\\
   0,&\hbox{if $\zeta\ne\eta$ otherwise,}\\
   -\sum_{\eta'\ne\eta}M_{\eta'\eta},&\hbox{if $\zeta=\eta$,}\end{cases}
 \ee
 Here $\eta^{ij}$ denotes the configuration obtained from $\eta$ by
exchanging the values of $\eta_i$ and $\eta_j$. $M^{(0)}$ is not invertible
but \eqref{Mdef} implies that it has a triangular structure so that
\eqref{tosolve} may be solved by back substitution.  We have used this
procedure to determine the coefficients $c_k$ of \eqref{coeffind} to order
$16$ and of the densities $\langle\eta_i\rangle$, $i=1,\ldots,5$, to order
$16-i$; the results are given in Appendix~\ref{sec:coeffs}.  The densities
behind the blockage may be obtained from this table and the symmetry
$\langle\eta_i\rangle=1-\langle\eta_{1-i}\rangle$.  Although
the $c_k$ given in the table alternate in sign we believe, on the basis of
an argument given below, that this alternation will break down at $k=17$.

In the next section we make, for the purposes of discussion,

 \smallskip\noindent
 {\bf Assumption A:}   The series
$\sum_{k=0}^\infty c_kr^k$ has a nonzero radius of convergence and
represents the current $j(r)$ for those $r\ge0$ for which it converges.

 \smallskip\noindent
 We ask what this assumption, together with the knowledge of the first
sixteen $c_k$, suggests about the behavior of $j(r)$.  Such a procedure is
clearly fraught with uncertainty, and any conclusions drawn must be
approached with skepticism, both because the series might diverge or, if
convergent, might represent $j(r)$ only in a limited region or not at all,
and because knowledge of any finite number of coefficients of a power
series can tell us nothing with surety of the behavior of the sum.
Nevertheless we feel that there is justification for our approach: the
analysis of the series seems fairly compelling, and the conclusions we
reach are consistent with the findings of \cite{JL2}.

\section{Analysis of the series for the current\label{sec:current}}

Let us define 
 \be
J(r)=\sum_{k=0}^{\infty} c_kr^k,\qquad 
  J_n(r)=\sum_{k=0}^n c_kr^k,\label{bjdef}
 \ee
 where the coefficients $c_k$ are defined by \eqref{coeffind} and
$c_0,\ldots,c_{16}$ are given in Table~\ref{table:current} of
Appendix~\ref{sec:coeffs}; as indicated above, we assume for the purpose of
this discussion that the series for $J(r)$ has a positive radius of
convergence.  We let $J(r)$ also denote the analytic continuation of the
sum of this series to other values of $r$.  The key tool in our analysis
will be the fact that for any function $f(z)$ analytic at $z=0$ the Taylor
series of $F(r):=f(J(r))$ in powers of $r$ can be calculated up to order 16
from the known coefficients $c_k$; we will write $F_{16}(r)$ for the
resulting polynomial.

We first note that for $k=1,\ldots,16$ the coefficients $c_k$ alternate in
sign and that for values of $k$ in the range $3\lesssim k\lesssim13$ the
ratio $c_k/c_{k-1}$ varies in magnitude in a reasonably narrow range around
$2/3$.  We accept the natural explanation of this behavior: that $J(r)$ has
a simple pole with relatively large residue at a point $r=r_0\approx-1.5$,
affecting the early coefficients of the series, but that this effect is
masked in later coefficients by weaker singularities closer to the origin.
Further analysis of the series will be aided by a more precise location of
$r_0$.  There are various methods to achieve this; we discuss two of them
in Appendix~\ref{sec:r0}.  The conclusion is that we may take $r_0=-1.5437$
and expect this value to be accurate within an error of $\pm 0.0005$.  In
the remainder of this section we will often study the power series for
$(r-r_0)J(r)$ (or some similarly constructed function---see, e.g.,
\eqref{Xrdef}) rather than that for $J(r)$, in order to eliminate the
strong effect of this pole and thus make other singularities of $J(r)$ more
apparent in the behavior of the power series.

\begin{figure}[ht!]
\begin{center}\includegraphics[width=10cm]{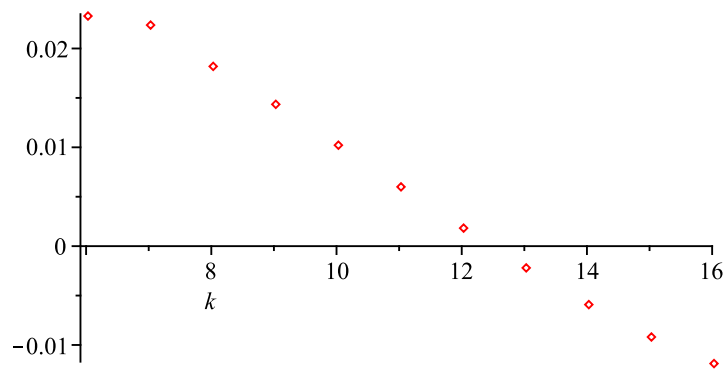} \end{center}
\caption{The Taylor coefficients of $(r-r_0)J(r)$.}
\label{fig:coeffj} \end{figure}

We now consider the question of whether or not $r_c=1$.  Under
Assumption~A, the radius of convergence of $J(r)$ can be at most $r_c$, so
that the Taylor coefficients of $J(r)$, or of $(r-r_0)J(r)$, must grow (for
large $k$) at least as fast as $r_c^{-k}$. \cfig{fig:coeffj} shows the latter
coefficients for $k=6,\ldots,16$; there is no sign of exponential growth,
suggesting that $r_c=1$.

\begin{center} \begin{figure}[ht!]
\begin{center}\includegraphics[width=10cm]{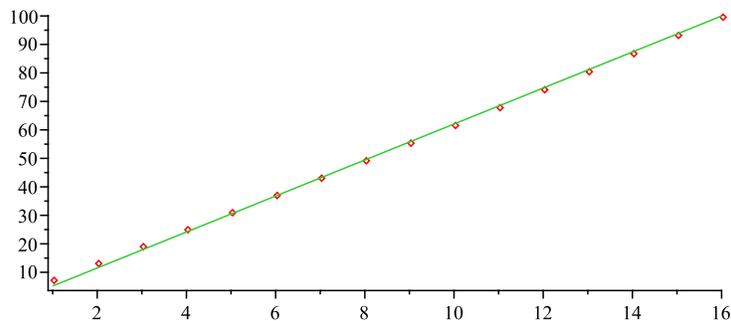} \end{center}
\caption{A plot of $\log^2u_k$ together with a straight line fit, where the
$u_k$ are the Taylor coefficients of $(1/4-J(r))^{-1}$.}
 \label{fig:sqrtk} \end{figure}\end{center}

Accepting provisionally that $r_c=1$ we now ask about the form of the
singularity in $J(r)$ there.  The heuristic formula \eqref{ljf} suggests
that 
 \be\label{ansatz}
J(r)\sim1/4-\gamma\, e^{-a/(1-r)}\quad \hbox{ for $r\lessapprox1$}, 
  \qquad \gamma,a>0,
 \ee
 so that $(1/4-J(r))^{-1}\sim\gamma^{-1}\exp(a/(1-r))$.  To verify the
consistency of this ansatz we observe (see Appendix~\ref{sec:asymp}) that
the Taylor coefficients of $\exp(a/(1-r))$ grow as
${\rm const}\cdot k^{-3/4}e^{2\sqrt{ak}}(1+o(1))$.  With $u_k$ the Taylor
coefficients of $(1/4-J(r))^{-1}$ we use the function Statistics[Fit] in
Maple to obtain a fit (also shown in \cfig{fig:sqrtk}) over the range
$7\le k\le16$ of the form $u_k\sim A_1\sqrt k+B_1\log k+C_1$ with
$A_1=2.82$, $B_1=-0.495$, and $C_1=0.116$; the residual standard deviation
in the fit, that is, the square root of the ratio of the residual sum of
squares to the number of degrees of freedom (here 7), is 0.0002, quite
small.  The value of $A_1$ is consistent with the value $a\approx2$ which
we discuss below; B1, C1 (and higher terms if sought) are more sensitive to
corrections to the exact type of the singularity, to other singularities,
etc.

To explore \eqref{ansatz} further we consider 
 \be
X(r)=\log[(r-r_0)(1/4-J(r))].\label{Xrdef}
 \ee
If \eqref{ansatz} holds then $X$ will have a simple pole at $r=1$, which by
itself would lead to a Taylor series for $X$ with all coefficients equal to
$-a$; due to the presence of other singularities we would of course expect
such constancy at best asymptotically.  A singularity of $X$ for $|r|<1$
would lead to exponentially growing coefficients, while other singularities
on the unit circle would modify the asymptotic form.

\begin{figure}[ht!]
\begin{center}\includegraphics[width=10cm]{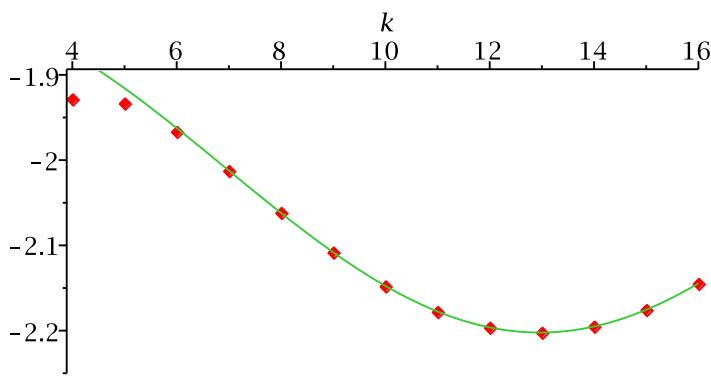} \end{center}
\caption{The Taylor coefficients of $\log[1/4-(r-r_0)J(r)]$, together with the
  curve $A+B\cos(C(k-D))$ with parameters given by \eqref{fitparams} .}
\label{fig:coeffx} \end{figure}

The Taylor coefficients $x_k$ of $X$ are plotted in \cfig{fig:coeffx}.
There is no sign of exponential growth, again suggesting that $r_c=1$ (and
more generally that $J(r)\ne1/4$ for $|r|<1$).  At a first approximation
one has $x_k\approx-2$, which is consistent with \eqref{ansatz} with
$a\approx2$.  Overall, we believe that this analysis gives good support for
the conclusion that $r_c=1$ and that $j(r)$ has a singularity of the
general form \eqref{ansatz} at $r=r_c=1$.

Somewhat more speculatively, one may in fact obtain a more detailed
hypothesis for the singularity structure of $J(r)$.
\cfig{fig:coeffx}  suggests an oscillatory behavior rather
than constancy for the $x_k$.  Using the function Statistics[Fit] in Maple
we obtain a very good fit (also shown in \cfig{fig:coeffx}) to a
function of the form $x(k)=A+B\cos(C(k-D))$, with
 \be\label{fitparams}
A=-2.00922,\quad B=0.193059,\quad C=0.260931,\quad D=0.919233,
 \ee
  over the range $7\le k\le16$.  The residual standard deviation is
0.00025, about one hundredth of one percent of the typical value being
fitted.

The function $\hat X(r)$ with Taylor series given exactly by
$\sum_{k=0}^\infty x(k)r^k$ is
 \be
  \hat X(r)
  =\frac{A}{1-r}+\frac{B[\cos CD-r\cos(C(D+1))]}{ 1-2r\cos C+r^2},
 \ee
 with simple poles at $r=1$ and $r=e^{\pm iC}$.  This suggests that one may
well-approximate $X(r)$ by the function $Y(r)$ whose Taylor series agrees
asymptotically with that of $\hat X(r)$ but has the
same first 16 terms as does that of $X(r)$,
 \be
 X(r)\approx Y(r):=\hat X(r)+\sum_{k=0}^{16}(x_k-x(k))r^k,
 \ee
 and correspondingly approximate 
 \be\label{Kdef}
J(r)\approx K(r):=\frac14-\frac{e^{Y(r)}}{r-r_0}.
 \ee
 By construction the first 16 coefficients in the power series for $K(r)$
agree with those of $J(r)$; the $17^{\rm th}$ coefficient is negative,
suggesting that the alternation of sign noted in Table~\ref{table:current}
breaks down at this point.

One must again treat this analysis with caution; certainly for example the
exact values of $A,\ldots,D$ depend on the range of indices $k$ over which
the fit is performed.
 
\begin{rem} One can attempt a similar analysis of the power series for the
densities $\rho_i(r)$ for small values of $i$, whose coefficients are also
given in Appendix B.  It is natural to guess that, near $r=1$, $\rho_i$
would have a decomposition similar to \eqref{ansatz}:
$\rho_i(r)=\rho^{\rm reg}_i(r)+\rho^{\rm sing}_i(r)$, with
$\rho^{\rm reg}_i$ regular at $r=1$ and $\rho^{\rm sing}_i$ having an
exponential singularity $e^{-A/(1-r)}$.  (Since the current is given by
$j(r)=r\langle\eta_0(1-\eta_1)\rangle=\langle\eta_i(1-\eta_{i+1})\rangle$,
$i\ne0$, where the expectation is taken in any NESS with maximal current,
occurrence of the exponential singularity of \eqref{ansatz} in the current
implies that it must occur in densities and/or in the two-body correlation
functions.)  The key to the analysis of the current, however, was that we
knew (or could hypothesize) that the regular part of the current at $r=1$
was precisely $1/4$, and could therefore isolate the singular part for
further study through the definition \eqref{Xrdef} of $X(r)$.  On the other
hand, while we know that $\rho_i(1)=\rho^{\rm reg}(1)=0.5$, it appears that
the derivatives of $\rho_i$ (or equivalently of $\rho^{\rm reg}_i$) at
$r=1$ are nonzero, and we have not been able either to determine reliably
any terms beyond the constant one in the expansion of $\rho^{\rm reg}_i$
around $r=1$ or to isolate $\rho^{\rm sing}_i$ so that its behavior could
be studied.  \end{rem}

\section{A solvable semi-infinite model \label{sec:semi}}

We would like to understand the convergence of the finite volume current
$\jring_L(r)$ (or $j_L^{\alpha,\beta}(r)$) as $L\to\infty$, with particular
attention to the behavior in the complex $r$ plane, and for this purpose
consider a semi-infinite system in which this behavior can be determined
completely, using the work of Derrida, Evans, Hakim, and Pasquier
\cite{DEHP}.  To obtain this model we begin with the open system as
described in Section~\ref{sec:intro}, with $\beta=1$, and then replace the
portion of this system lying to the left of the distinguished bond by a
reservoir of particles, so that the site immediately to the left of this
bond is always full.  Now we study the portion of the system to the right
of the bond; particles enter this system from the left at rate $r$ and (in
the finite system, now of $L$ sites) exit on the right at rate 1.  The
current in this system is an upper bound for that in the original model
\cite{JL2,Ligg2}.

The system may alternatively be viewed as a symmetrized TASEP on $\bZ$
\cite{PS}: the configurations are assumed to be symmetric under the joint
action of the reflection $i\to 1-i$ exchange of particles and holes, and
the dynamics are such that a transitions occur simultaneously on
symmetrically located bonds, at rate $r$ on the bond from 0 to 1, which is
invariant under the symmetry, and at rate 1 on all others.  Note that if
the symmetry condition is satisfied by the initial condition then it is
preserved by the dynamics.  In this language the model is again related to
a last passage percolation model, a discrete-time version of which was
studied in \cite{BR}.

  The current $\jsemi_L$ in this system is computed in \cite{DEHP}:
 \be\label{jsemi}
 \jsemi_L(r)=\frac{r \Qsemi_{L-1}(r)}{\Qsemi_L(r)},
 \ee
 where the polynomial $\Qsemi_L$ has order $L$ and is given explicitly by
 \be
\Qsemi_L(r) = \sum_{j=0}^L \frac{L+1-j}{L+1}\binom{L+j}{L}r^j,
 \ee
 and implicitly through the recursion
 \be\label{recur}
 \Qsemi_{L-1}(r)=(1-r)\Qsemi_L(r) + \frac{1}{L+1}\binom{2L}{L}r^{L+1},
 \quad  L\ge1,
 \ee
 with $\Qsemi_0(r)=1$.  It follows from \eqref{recur} that
 \be\label{jsemips}
 \jsemi_L(r)=r(1-r) + O(r^{L+2}).  
 \ee
  It is further shown in \cite{DEHP} that 
 \be\label{asymp}
  \jsemi(r)=\lim_{L\to\infty}\jsemi_L(r)
  =\begin{cases}
   r(1-r),& \hbox{if $0\le r<1/2$,}\\ 1/4,& \hbox{if $r\ge1/2.$}\end{cases}
 \ee
 Thus the critical value of $r$ in this model, $\rsemi_c=1/2$, is
very different from $r_c$, and moreover the behavior of $\jsemi(r)$ near
$\rsemi_c$ is very different from that conjectured in \eqref{ansatz} for
$j(r)$ near $r_c$.

\begin{figure}[ht!]
\begin{center}\includegraphics[width=10cm]{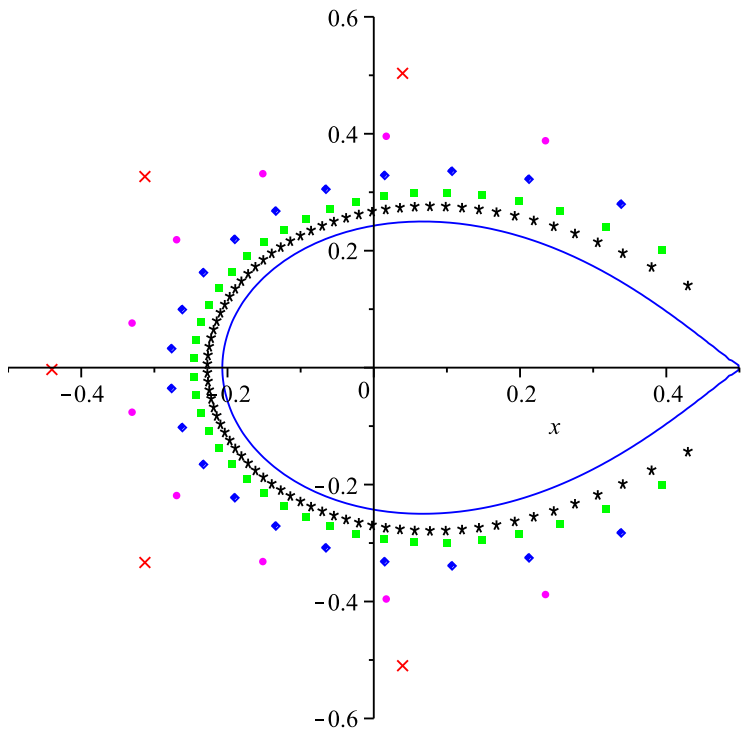} \end{center}
\caption{Zeros of the denominator $\Qsemi_L(r)$: The symbols are: $L=5$, red
cross, $L=10$, magenta circle; $L=20$; blue diamond; $L=40$, green square;
$L=80$, black star.  The solid line is the curve
$\Gamma=\{r\mid|r(1-r)|=1/4\}$.} \label{fig:goodpic} \end{figure}

The analysis which leads to \eqref{asymp} can be carried out also in the
complex $r$ plane.  The simple closed curve $\Gamma$ defined by
$|r(1-r)|=1/4$, ${\mathop{\rm Re}}\;r\le1/2$, divides the plane into an
interior region $\Omega_1$ and an exterior region $\Omega_2$; see
\cfig{fig:goodpic}.  Then for complex $r$,
 \be\label{conv}
   \jsemi(r)=\lim_{L\to\infty} \jsemi_L(r)
  =\begin{cases} r(1-r),& \hbox{if $r\in \Omega_1$,}\\
    1/4,&\hbox{if $r\in\Omega_2.$}\end{cases}
 \ee
 We see also that $|\jsemi(r)|$ is continuous on $\Gamma$. 

The mechanism for the convergence \eqref{conv} can be explored since the
zeros of the polynomial $\Qsemi_L$ can be obtained and plotted numerically
for reasonably large systems; this ``Lee-Yang'' approach to non-equilibrium
systems was introduced in \cite{A} and has been applied to the open TASEP
in \cite{BE}.  The zeros appear to fall on a smooth curve which
approaches $\Gamma$ as $L$ increases; the limiting curve $\Gamma$ and the
density of the zeros were computed in \cite{BE}.  The distance of the zeros
from $\Gamma$ decreases approximately as $1/L$, and the rightmost (complex
conjugate) zeros lie at a distance of order approximately $1/\sqrt{L}$ from
the point $r=1/2$.  See \cfig{fig:goodpic}, which plots the curve $\Gamma$
and the zeros of $\Qsemi_L$ for $L=5$, $10$, $20$, $40$, and $80$.  These
zeros are of course poles of $\jsemi_L$.  Note from \eqref{jsemips} that
there is no trace of these singularities in the first $L+1$ terms of the
power series for $\jsemi_L$; rather they are hidden in the $O(r^{L+2})$
remainder.

 \medskip\noindent
 {\bf Comparison of the models:} We now want to describe a certain
parallelism between the blockage and semi-infinite models, and to suggest
that this may extend to some properties which can be established for the
latter but not for the former. To make specific comparisons, we will
consider the finite-volume version of the blockage model to be that on the
ring, but similar conclusions hold for the interval.  

\begin{figure}[ht!]
\begin{center}\includegraphics[width=10cm]{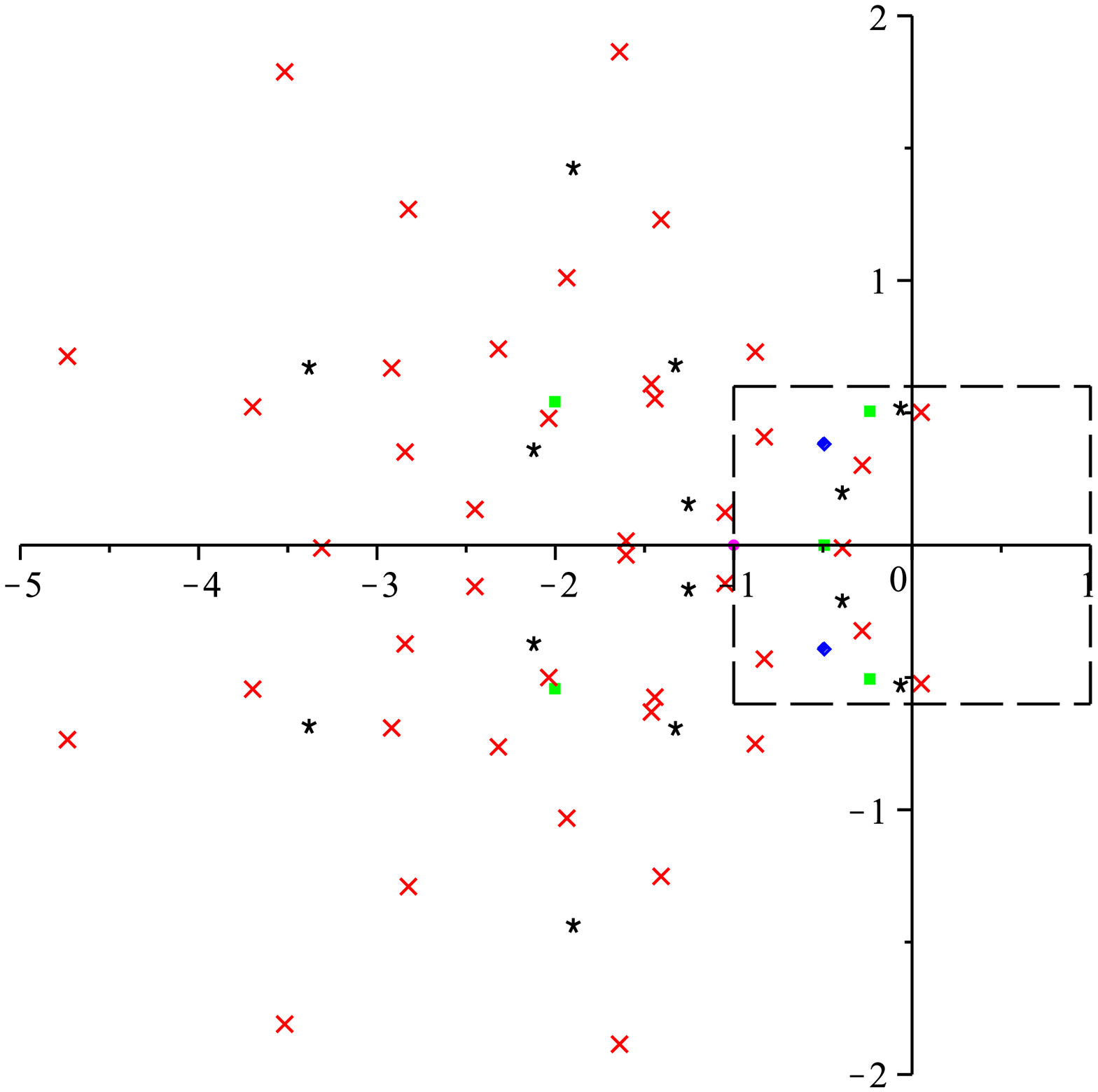} \end{center}
\caption{Zeros of the denominator $\Qring_L(r)$: $L=1$, magenta circle;
$L=2$; blue diamond; $L=3$, green square; $L=4$, black star; $L=5$, red
cross.  An expanded view of the dashed square is given in
\cfig{fig:simroots}(b).} \label{fig:roots15} \end{figure}

Before discussing this parallelism, however, we note a key difference
between the models: in the representation \eqref{jsemi} of $\jsemi_L$ the
numerator and denominator are polynomials in $r$ of degree $L$, while if
for the finite-volume blockage model we write
 \be
 \jring_L(r) = \frac{\Pring_L(r)}{\Qring_L(r)},
 \ee
 with $\Pring_L(r)$ and $\Qring_L(r)$ polynomials, then the degree of
$\Pring_L$ and $\Qring_L$ grows exponentially with $L$.  This of course
makes it difficult to carry out computations in the complex $r$ plane for
the blockage model similar to those described above for the semi-infinite
model.  To illustrate the difficulty we have calculated $\Qring_L(r)$ for
$L\le 5$; the degrees of these polynomials are $1$, $2$, $5$, $14$, and 42
for $L=1$, 2, 3, 4, and 5, respectively. Their zeros are shown in
\cfig{fig:roots15}, and one may compare the complexity of this figure with
the relative simplicity of \cfig{fig:goodpic}.

We next note  two particular similarities between the models.  

\begin{figure}[p!]
\begin{center}\includegraphics[width=11.2cm]{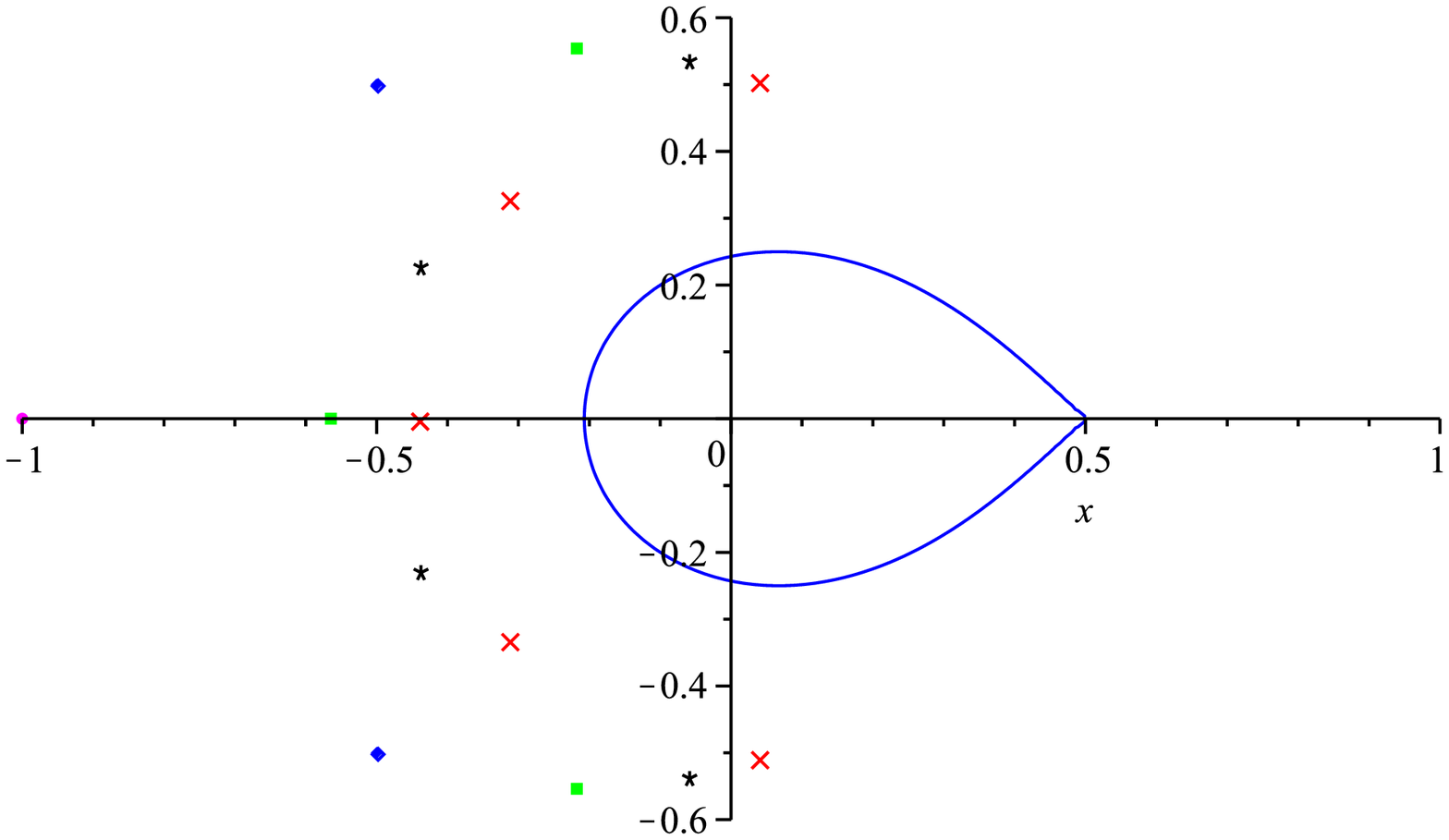} \end{center}
\begin{center} Figure \ref{fig:simroots}(a)\end{center}
\begin{center}\includegraphics[width=11.2cm]{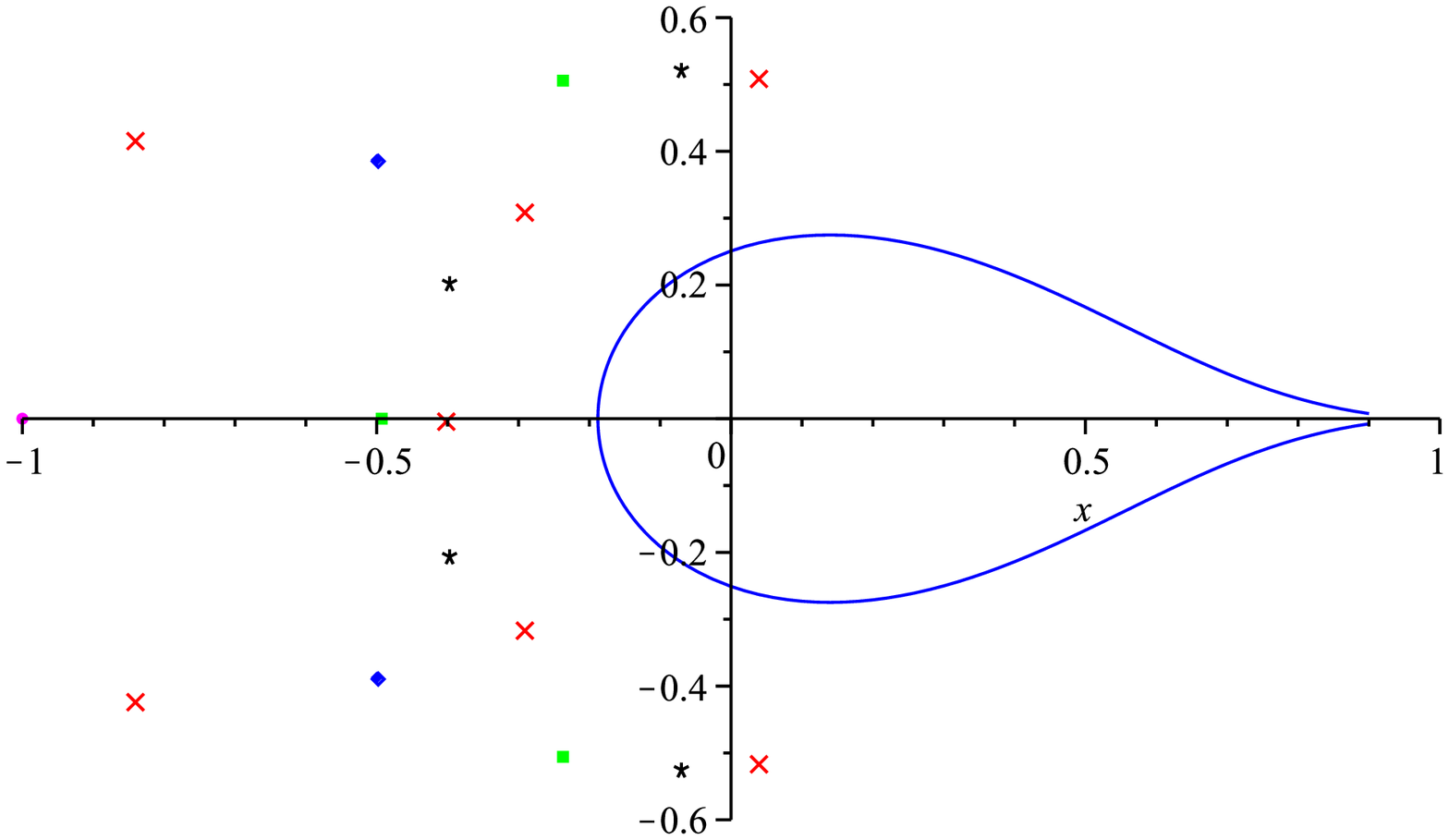} \end{center}
\begin{center} Figure \ref{fig:simroots}(b)\end{center}

\caption{Zeros of the denominator of the current for the semi-infinite
model (a) and the blockage system on the ring (b), for $N=1$, magenta circle;
$N=2$; blue diamond; $N=3$, green square; $N=4$, black star; $N=5$, red
cross.  Also shown are the limiting curve $\Gamma$ for the semi-infinite
model and an approximation to the  conjectured limiting curve $\Gammahat$
of \eqref{gammahat}.}
 \label{fig:simroots} \end{figure}

\begin{enumerate}

\item  In each model the coefficient of order $k$ of the Taylor series for
the finite volume current, $\jring_L(r)$ or $\jsemi_L(r)$, is
independent of $L$ for $L\ge k$; compare \eqref{jbps} and \eqref{jsemips}.
The parallelism is not complete: if we pass to the power series obtained
from the $L$-independent terms we obtain in the semi-infinite model just a
polynomial of degree two, $r(1-r)$, and in the original model the rather
complicated function $J(r)$ studied in Section~\ref{sec:ps}, but for our
present purposes we will ignore this distinction.

\item Although we have emphasized above that the singularity structure of
$\jring_L(r)$ is much more complicated than that of $\jsemi_L(r)$, there
nevertheless are similarities.  To see this, consider the two
plots of \cfig{fig:simroots}.  Here we show:

\begin{itemize}
 
\item for the semi-infinite model (\cfig{fig:simroots}(a)) the zeros of
$\Qsemi_L$ for $L=1,2,3,4,5$, as well as the limiting curve $\Gamma$;

\item for the blockage model (\cfig{fig:simroots}(b)), only the zeros of
$\Qring_L$, $L=1,2,3,4,5$, which lie near the origin; for the window
chosen, exactly $L$ zeros appear for $\Qring_L$ (as for the semi-infinite
model) except for $L=5$, for which two extra zeros, farthest from the
origin, also appear.  Also shown is a possible limiting curve discussed
below.

\end{itemize}

The similarity of these sets of zeros is clear.

\end{enumerate}

This parallelism suggests the following possible behavior for the blockage
model:

\begin{itemize}

\item Some subset of the zeros of $\Qring_L$ converge as $L\nearrow\infty$
  to a curve $\Gammahat$ somewhat similar to $\Gamma$.  The simplest
  hypothesis is that $\Gammahat$ is determined by the relation 
 \be\label{gammahat}
|J(r)|=\frac14,
 \ee
 in parallel with the definition of $\Gamma$.  The curve shown in
 \cfig{fig:simroots}(b)  was obtained by numerical solution of $K(r)=1/4$,
 where $K(r)$ is the approximation to $J(r)$ defined in \eqref{Kdef}. 

\item In the interior of the curve $\Gammahat$,
$J(r)$ is analytic and the finite volume current $\jring_L(r)$ converges to
$J(r)$ as $L\nearrow\infty$.  In particular, Assumption~A of
Section~\ref{sec:ps} is valid.

\item The behavior of $J(r)$ outside $\Gammahat$ has no direct connection
  with the behavior of the finite volume current $\jring_L(r)$ there.

\end{itemize}

\noindent
This suggested behavior is not sufficiently well founded to be called a
conjecture, but we hope that it may be a guide to further investigation.

\medskip
\noindent {\bf Acknowledgments:} We have greatly benefited from discussions
with M.~Bramson and B.~Derrida, and also thank M.~den~Nijs,
T.~Sepp\"al\"ainen, V.~Sidoravicius, and H.~Spohn for helpful conversations
and correspondence.  J.L.L and E.R.S. thank the IHES for hospitality.  The
work of OC was supported in part by NSF Grants DMS-0600369 and DMS-1108794,
and that of JLL in part by NSF Grant DMR-1104501 and AFOSR grant
FA9550-10-1-0131.

\appendix

\section{Agreement of $J_{\rm max}(r)$ and $j(r)$\label{cont}}

In this appendix we sketch a proof that the current $j(r)$ defined in
Section~\ref{sec:intro} is identical to the current $J_{\rm max}(r)$
introduced in \cite{S} and shown there to be continuous, showing that
$j(r)$ is continuous.  

Consider three versions of the 1-d TASEP process: $\eta_t$ is a process on
$\bbz$ with initial configuration $\eta_0(i)=1$ if $i\le0$, $\eta_0(i)=0$
if $i>0$; $\zeta_t$ is likewise a process on $\bbz$ but with arbitrary
initial configuration $\zeta_0$, and $\tau_t$ is a process on the interval
$\{-L+1,\ldots,L\}$ with entry and exit rates 1 and with initial
configuration the restriction of $\eta_0$ to this interval.  Transitions in
these processes occur at rate 1 on all bonds except $\{0,1\}$, where the
rate is $r$; moreover, the dynamics are coupled so that the same Poisson
clock on each bond controls transitions in each process, and the clocks on
$\{-L,-L+1\}$ and $\{L,L+1\}$ control also the entry and exit,
respectively, of particles in the $\tau$ process.

Suppose now that $\mu$ is a measure on $\{0,1\}^{\bbz}$ which is invariant
for the TASEP dynamics with the rates given above and has (maximal) current
$j(r)=r\langle\eta_0(1-\eta_1)\rangle_\mu$. (For the sake of precision we
define also $j^{1,1}_L(r)=r\langle\eta_0(1-\eta_1)\rangle_{\mu^{1,1}_L}$
with $\mu^{1,1}_L$ the invariant measure for the open system.)  Let
$N_\eta(t)$ be the number of particles that cross the special bond in the
$\eta$ process between times $0$ and $t$, with $N_\zeta(t)$ and $N_\tau(t)$
defined similarly.  From the coupling one sees easily that
$N_\tau(t)\ge N_\eta(t) \ge N_\zeta(t)$ for all $t$ and all $\zeta_0$, so
that for $t>0$,
 \be\label{ineq}
 \E{\frac{N_\tau(t)}t}\ge \E{\frac{N_\eta(t)}t}
   \ge \E{\frac{\langle N_\zeta(t)\rangle_\mu}t},
 \ee
 where $\E{\cdot}$ denotes the average over the dynamics and in
$\langle N_\zeta(t)\rangle_\mu$ the initial configuration $\zeta_0$ is
 averaged over $\mu$.  But also
 \begin{eqnarray}\label{currlim2}
  j(r)&=&\E{\frac{\langle N_\zeta(t)\rangle_\mu}t},\quad t>0,\\
  \label{currlim3}
  j^{1,1}_L(r)&=&\lim_{t\to\infty}\E{\frac{N_\tau(t)}t},
 \end{eqnarray}
 We show below that $\lim_{t\to\infty}\E{N_\eta(t)/t}=J_{\rm max}(r)$; with
\eqref{currlim} and (\ref{ineq})--(\ref{currlim3}) this completes the proof.
 
For $n\ge1$ let $T_n$ be the time of the $n^{\rm th}$ crossing of the
$\{0,1\}$ bond by a particle in the $\eta_t$ process.  It is shown in
\cite{S} that $\kappa(r)=\lim_{n\to\infty}T_n/n$ exists almost surely and
is constant.  The current $J_{\rm max}$ is defined to be $1/\kappa(r)$,
so that $J_{\rm max}(r)=\lim_{n\to\infty}n/T_n$ a.s.  It is easy to see
that $N_\eta(t)\to\infty$ and so  $T_{N_\eta(t)}\to\infty$ a.s. as
$t\to\infty$; from this and the inequalities
 \be\label{basic}
 \frac{N_\eta(t)+1}{T_{N_\eta(t)+1}}-\frac1{T_{N_\eta(t)+1}}
  <\frac{N_\eta(t)}t
  \le\frac{N_\eta(t)}{T_{N_\eta(t)}},
 \ee
we see that $\lim_{t\to\infty}N_\eta(t)/t=J_{\rm max}$ a.s. We complete the
proof by verifying that the family of random variables
$\{N_\eta(t)/t\mid t\ge1\}$ is uniformly integrable, that is, that
 \be\label{uc}
 \lim_{c\nearrow\infty}\,\max_{t\ge1}
   \,\bE\left[\frac{N_\eta(t)}{t}
   \chi_{\bigl\{\frac{N_\eta(t)}{t}\ge c\bigr\}}\right],
 \ee
 where $\chi_S$ denotes the characteristic function of the set $S$.

 Note first that $T_n\ge S_n$, where $S_n$ is the sum of $n$ independent
exponential random variables of mean 1 (these can be taken, for example, to
be the variables $Y_{i,i}$, $1\le i\le n$, of \cite{S}).  Then for $c\ge2$,
and using  $T_{N_\eta(t)}\le t$, we have
 \begin{eqnarray}\label{est1}
\bE\left[\frac{N_\eta(t)}{t}
   \chi_{\{N_\eta(t)\ge ct\}}\right]
 &=&\sum_{n=2}^\infty \bE\left[\frac{n}{t}\chi_{\{N_\eta(t)=n\}}
   \chi_{\{n\ge ct\}}\right]\nonumber\\
 &\le&\sum_{n=2}^\infty \bE\left[\frac{n}{S_n}
   \chi_{\{n\ge cS_n\}}\right].
 \end{eqnarray}
 Now $n/S_n$ is gamma-distributed with density $n^nx^{n-1}e^{-nx}/(n-1)!$,
 so that, using $n!\ge n^ne^{-n}$, we have
 \be\label{est2}
\bE\left[\frac{n}{S_n}
   \chi_{\{n\ge cS_n\}}\right]
  \le\frac{n^n}{(n-1)!}\int_0^{1/c}x^n\,dx
  \le\frac{e^n}{c^{n+1}},
 \ee
 and \eqref{uc} follows from \eqref{est1} and \eqref{est2}.

\section{Tables of coefficients\label{sec:coeffs}}

We give in Table~\ref{table:current} below the coefficients in the power
series \eqref{bjdef} for the current $J(r)$, to order 16, and in
Table~\ref{table:densities} the coefficients in the series
 \be\label{psdens}
\rho_i(r) =\sum_{k=0}^\infty d_{ik}r^k
 \ee
for $\rho_i$, up to order $16-i$, for $i=1,\ldots,5$.  In
Table~\ref{table:current} we report the values to 20 decimal places. We
believe that these digits are reliable because in our computation of
$c_0,\ldots c_{15}$ the first 24 digits obtained were stable under increase
in the system size (this test is not available for $c_{16}$).  For 
$c_0,\ldots c_{13}$ a check for stability under change in the
precision used in the program found the same reliability.

 \begin{table}[ht!]  \begin{center}
\begin{tabular}{|c|r|c|c|r|}
\hline
\rule[-2mm]{0pt}{6mm} $k$&\multicolumn{1}{c|}{$c_k$}&&
   $k$&\multicolumn{1}{c|}{$c_k$}\\
\hline\hline
0  &  0.00000000000000000000&& 9  &  0.09696668201649961665\\
1  &  1.00000000000000000000&& 10 & -0.05607405058503243547\\
2  & -1.50000000000000000000&& 11 &  0.04033294103506195933\\
3  &  1.18750000000000000000&& 12 & -0.02482314806701423240\\
4  & -0.77889901620370370370&& 13 &  0.01477809455732788252\\
5  &  0.52961027553406380931&& 14 & -0.01328263357536883488\\
6  & -0.32787247554422253211&& 15 &  0.00277198065020739725\\
7  &  0.22700745336484005616&& 16 & -0.00936905520626202337\\
8  & -0.13514784111152134747&&&\\
\hline
\end{tabular}
\medskip
\caption{The coefficients of the power series $J(r)$.
\label{table:current}} 
\end{center}
 \end{table}

\begin{table}[ht!]
\advance\tabcolsep by -0.5pt
\begin{center}
\begin{tabular}{|c|r|r|r|r|r|}
\hline
\rule[-2mm]{0pt}{6mm} $k$&
  \multicolumn{1}{c|}{$d_{1k}$}&
  \multicolumn{1}{c|}{$d_{2k}$}&
  \multicolumn{1}{c|}{$d_{3k}$}&
  \multicolumn{1}{c|}{$d_{4k}$}&
  \multicolumn{1}{c|}{$d_{5k}$}\\
\hline\hline
 0 &0.00000000 & 0.00000000 & 0.00000000 & 0.00000000 & 0.00000000 \\
 1 &1.00000000 & 1.00000000 & 1.00000000 & 1.00000000 & 1.00000000 \\
 2 &-0.75000000 & -0.68750000 & -0.65625000 & -0.63671875 & -0.62304688 \\
 3 &0.45312500 & 0.40075231 & 0.37939743 & 0.36763019 & 0.36002612 \\
 4 &-0.32345016 & -0.31775049 & -0.32035731 & -0.32441081 & -0.32884964 \\
 5 &0.19113754 & 0.16723143 & 0.15934419 & 0.15807536 & 0.16042006 \\
 6 &-0.13508488 & -0.13527025 & -0.14002291 & -0.14544675 & -0.15090756 \\
 7 &0.08218276 & 0.07251374 & 0.06758191 & 0.06440565 & 0.06221118 \\
 8 &-0.05412219 & -0.05222456 & -0.05418729 & -0.05745286 & -0.06108373 \\
 9 &0.03743811 & 0.03658774 & 0.03602396 & 0.03505521 & 0.03382992 \\
 10 &-0.01961547 & -0.01492080 & -0.01314674 & -0.01296838 & -0.01376695 \\
 11 &0.01873543 & 0.02252206 & 0.02556734 & 0.02764994 & 0.02887908 \\
 12 &-0.00544623 & 0.00051738 & 0.00480082 & 0.00795647 &\\
 13 &0.01025427 & 0.01536259 & 0.02003591 &&\\
 14 &-0.00040069 & 0.00489313 &&&\\
 15 &0.00558944 &&&&\\
\hline
\end{tabular}
\medskip
\caption{The coefficients $d_{ik}$ of the power series
$\rho_i=\sum_kd_{ik}r^k$.   \label{table:densities}}
\end{center}
\end{table}

\section{The value of $r_0$\label{sec:r0}}

As observed in Section~\ref{sec:current}, the power series for the current
suggests the existence of a pole at some value $r_0\approx-1.5$.  In this
appendix we attempt to determine an accurate value for $r_0$ by two
independent methods.

\begin{figure}[ht!]
\begin{center}\includegraphics[width=10cm]{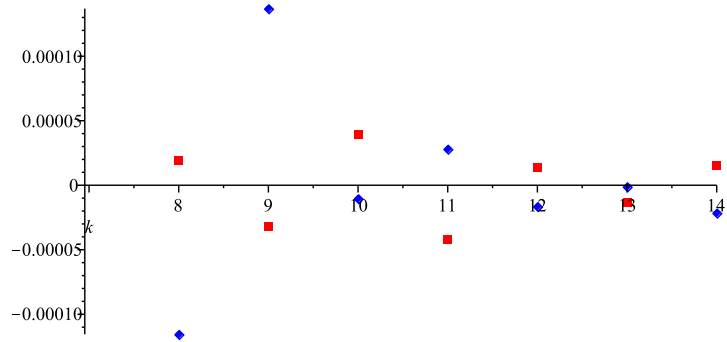} \end{center}
\caption{The difference $\hat w_k-\tilde w(k)$ for $r_0=3.5435$ (red
squares) and $r_0=3.5440$ (blue diamonds).}
\label{fig:wkhat} \end{figure}

 \smallskip\noindent
 {\bf Method 1.} We let $W(r)=(r-r_0)\,J(r)$ and compute, for various
values of $r_0$, the Taylor polynomial $W_{16}(r)=\sum_{k=0}^{16}w_kr^k$.
The oscillation in the coefficients $w_k$ should be minimized for the
correct value of $r_0$; to emphasize the oscillation we compute the second
difference $\hat w_k=w_{k+2}-2w_{k+1}+w_k$ as well as
$\tilde w(k)$, the least-squares linear fit to $\hat w_k$ over the range
$k=8\ldots14$,
and plot in \cfig{fig:wkhat} the difference $\hat w_k-\tilde w(k)$
for $r_0=-1.5435$ and $-1.5440$. The oscillations clearly change sign over
this range of $r_0$ and will be minimized at some intermediate value.

 \smallskip\noindent
 {\bf Method 2.} Consider the function $V(r)=1/(1/4-J(r))$, which has a
zero at $r=r_0$ and for which we expect a singularity at $r=1$, due both to
the vanishing of the denominator and to the (hypothetical) exponential
singularity of $J(r)$ suggested in \eqref{ljf}.  The fractional linear
transformation
 \be\label{flt}
u=u(r)=\left(\frac{1-r_1}{r_1}\right)\frac r{1-r},\quad\Leftrightarrow\quad
 r=r(u)=\frac{r_1u}{1+r_1(1-u)},
 \ee
 with $r_1=-1.5$, leaves the origin unchanged, carries $r=r_1$ to $u=1$,
and moves the singularity at $r=1$ as far away as possible, to $u=\infty$.
We write $\hat V(u)=V(r(u))$ and find that $\hat V_{16}(u)$ has one zero
$u_0$ very near to $u=1$, corresponding to $r_0=r(u_0)=1.5437$, and that all
other zeros are a distance at least 1.39 from $u=0$.  The value of $r_0$
found by this procedure is independent, to very high precision, of the
choice of $r_1$ in \eqref{flt} for $-3\le r_1\le-1$.

 \smallskip\noindent
 Thus the two methods above give good agreement and suggest that
$r_0\in[-1.540,-1.535]$.  We take $r_0=-1.5437$ and expect this
value to be accurate within an error of $\pm 0.0005$.

\section{Asymptotics of Taylor coefficients\label{sec:asymp}}

 In this section we find the large $k$ asymptotic behavior of the
coefficients $b_k$ in the Maclaurin series 
\begin{equation} \label{eq:st}
e^{a/(1-r)}=\sum_{k=0}^{\infty} b_k r^k,\ \ |r|<1,
 \end{equation}
where $a$ is a positive number. We have
 \begin{equation}
  \label{eq:exprck}
  b_{k}=\frac{1}{2\pi i}\oint_{\mathcal{C}_1}\frac{e^{a/(1-s)}}{s^{k+1}}ds
   =\oint_{\mathcal{C}_2} e^{-a/t-(k+1)\ln(1+t)}dt,
 \end{equation}
with $\mathcal{C}_1$  a circle of radius $R<1$ centered at $0$,
traversed anticlockwise, and $\mathcal{C}_2=-1+\mathcal{C}_1$.  The saddle
points of $-a/t-(k+1)\ln(1+t)$ are:
 \begin{equation}
  \label{eq:sad1}
 t_+= \frac{a+\sqrt{a(a+4k+4)}}{2(k+1)}\ \text{and}\ 
   t_-=  \frac{a-\sqrt{a(a+4k+4)}}{2(k+1)}.
 \end{equation}
The steepest variation path homotopic to $\mathcal{C}_2$ and passing
through $t_-$ is an analytic curve orthogonal at $t_-$ to the real line.
The saddle point method applied at $t=t_-$ gives
 \begin{equation}
  \label{eq:asympt}
  b_{k}={\rm const}\cdot k^{-3/4}e^{2\sqrt{ak}}(1+o(1)) \ 
  \text{as}\ k\to \infty.
 \end{equation}

\end{document}